# High efficiency end-fire 3-D optical phased array based on multi-layers $Si_3N_4/SiO_2$ platform

DACHUAN WU,[1] YASHA YI,[1, 2, *] AND YUXIAO ZHANG,[1]

[1]*Integrated Nano Optoelectronics Laboratory, Department of Electrical and Computer Engineering, University of Michigan, 4901 Evergreen Rd., Dearborn, Michigan 48128, USA*
[2]*Energy Institute, University of Michigan, 2301 Bonisteel Blvd., Ann Arbor, Michigan 48109, USA*
*\*yashayi@umich.edu*

**Abstract:** Beam steering device such as optical phased array (OPA) is a key component in applications of solid-state Lidar and wireless communication. The traditional single-layer optical phased array (OPA) results in a significant energy loss due to the substrate leakage caused by the downward coupling from the grating coupler structure. In this work we have investigated a structure based on multi-layers $Si_3N_4/SiO_2$ platform that can form a 3-D OPA to emit the light from the edge of the device with high efficiency, a 2-D converged out-coupling beam will be end-fired to the air. The high efficiency and wide horizontal beam steering are demonstrated numerically, the influence of vertical cross-talk, the delay length, number of waveguide layers, and the fabrication feasibility are also discussed.

## 1. Introduction

With the emerging applications such as solid-state Lidar (light detection and ranging), the beam steering based on integrated optical phased array (OPA) has drawn a lot of research efforts in the past decade [1]. Significant progress has been made including thermal tuning [2], electro-optics tuning [3], high sensitive wavelength tuning [4, 5], integrated on-chip light source [6], side lobe suppression by aperiodic or apodized array placement [7-9], etc.

The traditional way in on-chip integrated photonic research usually utilizes the *single* waveguide layer structure, which is also the case of most studies on OPA. For example, in [1-8], the device structure could be sophisticated due to various requirements, while all containing only one waveguide layer, and because of this, the OPA formed by a single layer can only have the exit beam upward. This actually is the reason of the relatively low emitting efficiency. When the OPA is placed in an environment that both its front and back side are uniform medium, the interference of light forms a beam not only to its front side, but also to its back side. In our previous work [10], we showed that a portion larger than 50% of light can be emitted to the substrate when an OPA is working in the case that its front side is air and back side is glass. However, as one of the main potential applications of integrated beam steering devices, the solid-state Lidar usually requires a detection range over at least 100 m. The light emitting efficiency of the beam steering devices, despite the development of light source and detector, is directly related to detection range of Lidar.

Several works have been attempted to address the relatively low efficiency challenge [11-16]. In [11], a structure configuration to emit light from the edge of the chip is utilized. An ultra-converged beam is also achieved in [12]. Further works aiming to confine the waveguide spacing to half-wavelength have been done by various approaches [13, 14]. However, these four works also employ the configuration of single waveguide layer. This does offer the convenience on tuning the phase of each waveguide [11, 13, 14], but the beam emitted by such a configuration is indeed a fan-beam, as the single waveguide layer can only form a 1-D OPA on the edge of the chip. The possibility of emitting a 2-D converged beam from the edge (end-fire) requires a 2-D OPA on the edge side. This is discussed in [15] and [16]. In [15], the performance of a 2-D end-fire OPA is numerically discussed, and a method utilizing nanomembrane transfer printing to fabricate a multi-layer structure with the top Si layer from SOI wafer is proposed and experimentally proved. In [16], a direct writing method based on ultrafast laser inscription (ULI) is applied to achieve a structure for the conversion between single-layer waveguides and 3-D waveguides, therefore, a 2-D OPA can be formed on the edge side.

In this work, we have studied a 3-D structure configuration based on multi-layer $Si_3N_4/SiO_2$ platform to achieve a 2-D convergent beam emitted from the edge. We numerically demonstrate the performance of this structure and discuss the main improvement on the energy efficiency in both the light input end and emitting end. The influence of vertical cross-talk, the engineering of delay length and the number of waveguide layers are also investigated.

## 2. Structure Configuration

The structure configuration is as shown in Fig. 1. The device consists of 6 $Si_3N_4$ layers with a thickness of 800nm, 5 $SiO_2$ layers with 500nm thickness are sandwiched by the $Si_3N_4$ layers, which is shown as Fig 1 (b). Each $Si_3N_4$ layer is patterned, Fig. 1 (c) is the top view of each $Si_3N_4$ layers, 9 waveguides with 800nm width are placed, the spacing between the center of each waveguide is $2\mu m$. In the red circled part, the length of the waveguides is gradually increased by a step of 6200nm.

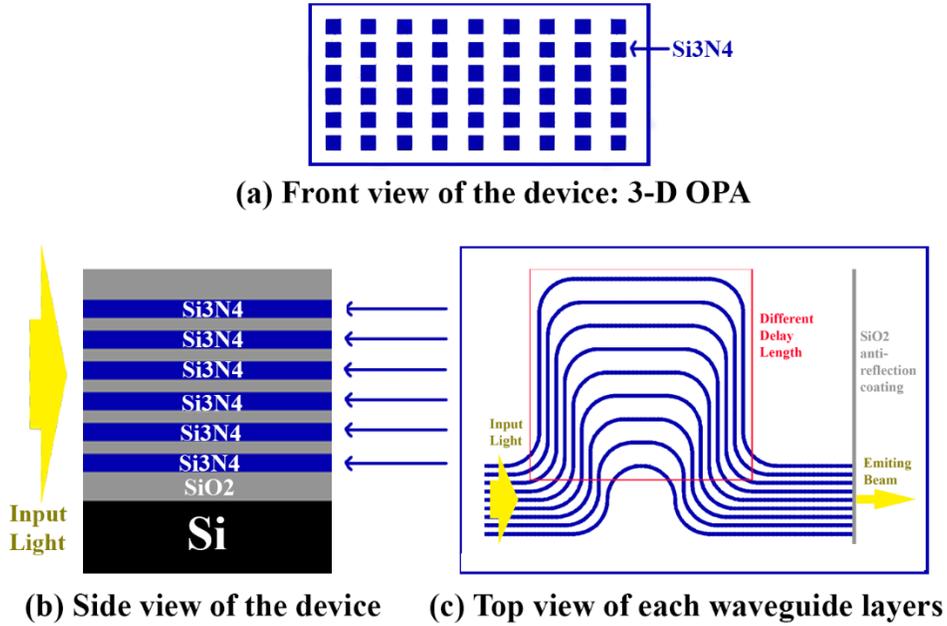

Fig. 1. Illustration of the structure. (a) Front view: the 3-D OPA is formed on the front edge of the device, (b) Side view: cross-section of the device, 6 $Si_3N_4$ layers of 800nm thickness and 5 $SiO_2$ layers of 500nm thickness, (c) Top view: pattern of each waveguide $Si_3N_4$ layers, contains 9 waveguides with 800nm width, spacing $2\mu m$.

The fabrication strategy of this device is discussed as follows. This structure can be fabricated on Si substrate. Firstly, a $SiO_2$ layer of $2\mu m$ is deposited as low index substrate; secondly, 6 patterned $Si_3N_4$ layers and 5 un-patterned $SiO_2$ layers are fabricated. The 6 patterned $Si_3N_4$ layers have exactly the same pattern, which is as shown in Fig. 1. There are two possible ways to fabricate this multi-layer structure. First, the method proposed in [12] can be utilized in this fabrication, the challenge in this method will be the control on the planarization step, this step will influence the thickness of the sandwiched $SiO_2$ layer, while the precise control of this thickness is crucial in this structure. Second, thanks to the identical patterns on each $Si_3N_4$ layer, it is possible to utilize the self-aligning method proposed in [15]. The fact that $Si_3N_4$ and $SiO_2$ have a low etching selectivity between each but both have a high selectivity against Si also can help on this method: the selectivity difference offers a possibility to etch down multiple layers of $Si_3N_4$ and $SiO_2$ within the same etching step where a Si is used as the mask. However, the extremely high aspect ratio will be an obstacle here. It is more realistic to fabricate this structure with a combination of the two methods: use the self-aligning method to etch down 2-3 $Si_3N_4$ layers within one step, and then use the multi-step process proposed in [12] to get the 6 patterned $Si_3N_4$ layers eventually. As long as the self-aligning method can handle more than one $Si_3N_4$ layer in one step, (the aspect ratio of the hole in the etching step will be 1.2 $\mu m$ /2.1 $\mu m$=0.57 in the case of etching down two $Si_3N_4$ layer and one sandwiched $SiO_2$ layer at once), the number of the total steps required will be significantly reduced.

After the fabrication of 6 patterned $Si_3N_4$ layers and 5 un-patterned $SiO_2$ layers, a final passivation $SiO_2$ layer needs to be deposited. Following this, the wafer will be diced, and the edge of the die will be polished to ensure the input coupling from the external laser source and the output coupling at the emitting end. As the final step, a quarter-wavelength $SiO_2$ layer (T=1550/(4*1.45)=267nm) will be deposited on the edge side, which can perform the antireflection function between $Si_3N_4$ waveguides and the air. The whole fabrication process follows the standard microchip fabrication process, so the method is CMOS compatible.

In this work, we only consider 9 waveguides in each layer, and the total periodicity of the array at the emitting end is $2\mu m$. It's worth to note that even though only 9 waveguides are considered in this work, it is possible to have more in each layer by the using of beam splitter tree [1]. Then, an Ω shape delay length structure is employed to create phase different between the waveguides. The difference in the length between each waveguides is the same, and the $2\mu m$ periodicity can eliminate the cross-talk between waveguides, therefore, the phase difference between each array at the emitting end will be same, which can satisfy the phased array condition in Equ. (1) in each waveguide layers (horizontal direction). In the vertical direction, because of the structure in each waveguide layers are exactly the same, the phase difference between each layer is 0, so this also satisfies the phased array condition between layers (vertical direction). So, this structure is able to emit a beam with 2-D convergence from the edge of the device.

In this work, we only consider 9 waveguides in each layer, and the spacing of the array at the emitting end is $2\mu m$. It's worth to note that even though only 9 waveguides are considered in this work, it is possible to have more waveguides in each layer by the using of a beam splitter tree [1]. The convergence of the light can be enhanced by a large number of array elements, so even though the convergence data presented in this work are obtained from 9 waveguides, it actually is possible to be further improved in principle. Then, an Ω shape delay length structure is employed to create a uniform phase difference between the waveguides, and the $2\mu m$ spacing can eliminate the cross-talk between waveguides. Therefore, the phase difference between each array at the emitting end will be same, which can satisfy the phased array condition in each waveguide layer (horizontal direction). In the vertical direction, because of the structure in each waveguide layers are exactly the same, the phase difference between each layer is 0, so it also satisfies the phased array condition between layers (vertical direction). As a total result, this structure is able to emit a beam with 2-D convergence from the edge of the device. Equ. 1 shows the phase condition:

$$sin\theta = \frac{\lambda_0 \cdot \varphi}{2\pi \cdot d}, \quad (1)$$

where $\theta$ is the emitting angle, $\lambda_0$ is the vacuum wavelength, $\varphi$ and $d$ are the phase difference and spacing between each array element.

The most important improvement of this structure will be the very high energy efficiency. First, in most of the previous studies, an external laser with single mode fiber is considered as the light source, however, the light will suffer a considerable loss at the input coupling no matter a vertical couple or a butt couple is employed, especially in the butt couple, the loss is usually significant because the thickness of the waveguide layer is usually ten times smaller than the mode field diameter (MFD) of a single mode fiber with a common core diameter of $8.2\mu m$. In this multi-layer structure, 6 $Si_3N_4$ layers and 5 sandwiched $SiO_2$ layers occupy $7.3\mu m$ vertically, thus, when the light is coupled from the single mode fiber to the on-chip waveguides, the coupling efficiency will be much higher than using single waveguide layer structures. A spot size converter using the similar coupling mechanism was proposed in [17], it first couples the light from a single mode fiber to a tapered stack of $Si_3N_4/SiO_2$ layers with similar size, then converts the spot shape to be a vertical line after the taper, and eventually couples the light to a Si waveguide with a much smaller size than the fiber. Based on their result, the mode overlap between the single mode fiber and the multi-layer spot size converter can be as high as 94%-99%. In that work, the thickness of $Si_3N_4$ layers is 225nm, while in this work, the thickness of the $Si_3N_4$ waveguide layers is 800nm, so the coupling efficiency may not be as high as their result, but we believe it is still obvious that the multiple $Si_3N_4$ layers can contribute to high input coupling efficiency.

On the other hand, this multi $Si_3N_4$ layer structure can also help with the apodized field distribution. In [9], a Gaussian-apodized phased array is utilized to suppress the side lobe, in that work, the apodized field distribution across the array is purposely designed. In this work, because of the coupling mechanism, the apodized field distribution across each $Si_3N_4$ waveguide layers is automatically formed. This is illustrated in Fig. 2.

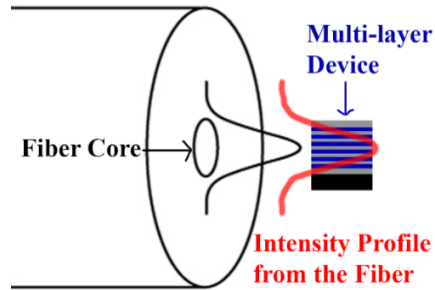

Fig. 2. Illustration of the apodized intensity profile of the input coupling, the total thickness of the device is equivalent to the MFD of a common single mode fiber

The high efficiency is also contributed by the emitting end. From Fig. 1, it can be seen that the OPA is formed on the edge of the device, the front side of the OPA is the antireflection coating and air, which is a uniform medium, and an out-coupling beam can be generated by the interference between each array element. In addition, because the beam is end-fired to the air, the backward emitting is highly suppressed. From the simulation result, the out-coupling efficiency at the emitting end can be as high as 82%, the results will be discussed in part 3.

## 3. Result and Discussion

In this work, the FDTD (finite difference time domain) method is utilized to simulate the structure. In the simulation, the model is set exactly same as Fig. 1, and a TM-polarized Gaussian pulse is applied as the light source, the pulse illuminates every waveguide. Wavelength range is set to be 1400nm to 1700nm, covering the 100nm wavelength tuning range which is used in [1]. Fig. 3 shows the farfield pattern of the device at 1550nm. It shows that the device generates a clear main lobe at $-1.53°$ horizontally and $-6.99E-4°$ vertically, this lobe has a horizontal FWHM (full width at half maximum) of $4.43°$ and a vertical FWHM of $10.96°$; two side lobes in horizontal direction can be observed at $-52.78°$ and $-47.77°$. The formation of the side lobes is a result of the large spacing between waveguides in each layer, while because of the periodicity ($2\mu m$) is not too much larger than the wavelength (1550nm), the side lobes are far away from the main lobe and not strong. On the other hand, the periodicity in vertical, which is the spacing between center of each waveguide layers is $1.3\mu m$, so only one clear main lobe is generated in vertical direction. The vertical FWHM is larger than horizontal FWHM, this is because the OPA covers only $7.3\mu m$ vertically, but $18.8\mu m$ horizontally.

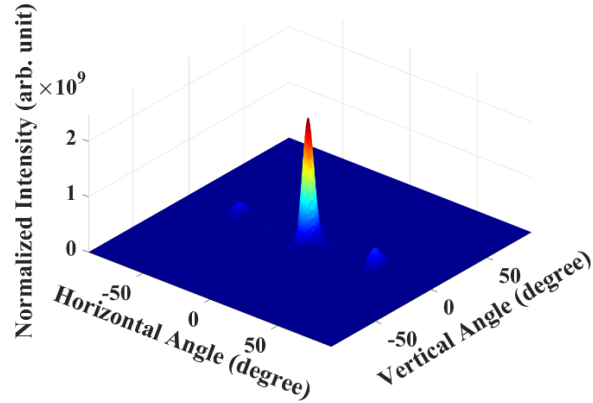

Fig. 3. Farfield pattern of the device at 1550nm, a clear 2-D converged beam is emitted by the device.

The wavelength tuning performance of the device is shown in Fig. 4. Fig. 4 (a) is the horizontal farfield contour map, it shows that the main lobe can be steered by wavelength tuning, the beam is steered from $10.99°$ at 1500nm to $-13.79°$ at 1600nm, a $24.78°$ steering range in 100nm wavelength range is achieved. This steering capability is achieved by the delay length structure, the length difference between each waveguide is fixed, so the phase difference between each array element in horizontal direction can be coherently changed by wavelength tuning. Fig. 4 (b) is the vertical farfield contour map, there is no length difference between waveguides in different layers, so the phase difference between each array element in vertical direction is always 0 in wavelength tuning, so the vertical farfield angle changes only slightly in the whole range, it only changes $3.13E-3°$. Fig. 4 (c) shows a comparison between the horizontal and vertical farfield angle. The beam steers linearly in horizontal direction with the wavelength variation, and the steering is ignorable in vertical direction. Fig. 4 (d) shows the variation of horizontal and vertical FWHM, the convergence of the beam keeps in the whole wavelength tuning range, the horizontal FWHM varies less than 7.67%, and the vertical FWHM varies less than 7.12%. Fig. 4 (e) shows the coupling efficiency in the tuning range, this efficiency is calculated by using the total energy emitted divided by the energy in all the waveguides right before the emitting OPA. The efficiency at the whole range is higher than 76.43%, this minimum value appears at wavelength of 1600nm, a maximum value of 82.22% can be observed at wavelength of 1570nm. Because of the main lobe dominates in the whole range, the emitting efficiency of the main lobe is close to the efficiency value in Fig. 4 (e). As discussed in part 2, the high efficiency is contributed by both the end-fire mechanism and the $SiO_2$ antireflection coating.

The tuning function with purely horizontal steering is achieved by the wavelength tuning only. In this device, it is not required to have a precise control of the phase in each waveguide, and hence, the number of degree of freedom required for operation is reduced from N (N is the number of waveguides in each layer) to 1, the operation principle is high simplified. In the application of Lidar, the traditional mechanical Lidar rotates the whole device to achieve the horizontal field of view (FOV), and the vertical FOV is achieved by utilizing multiple beam lines vertically, this requires each beam maintains its vertical angle during the rotation. This device can emit a 2-D converged beam that can be steered purely horizontally, so it is possible to utilize multiple devices together to build a multi-line solid-state Lidar.

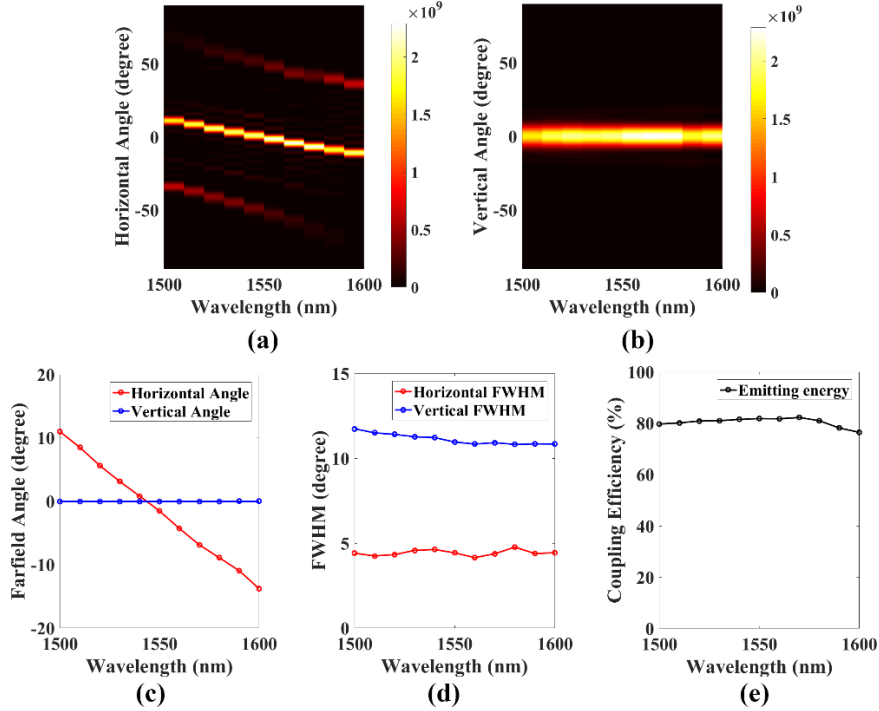

Fig. 4. Simulation result of the structure in Fig. 1. (a) Horizontal farfield contour map, a clear main lobe steers 24.78°/100nm, two side lobes can be observed, (b) Vertical farfield contour map, only one main lobe exists, no steering vertically (c) Comparison between the horizontal and vertical angle, (d) Comparison between the horizontal and vertical FWHM, (e) Coupling efficiency of the total energy emitted.

### 3.1 Influence of vertical cross-talk

The spacing between each waveguide in horizontal is selected to be $2\mu m$ to eliminate the cross-talk, this is a consideration of the fact that the phase of light in each horizontal waveguides are different. On the other hand, the thickness of sandwiched $SiO_2$ are set to be 500nm. Indeed, this thickness cannot fully eliminate the cross-talk between the waveguides in different layers. However, thanks to the exact same pattern in each waveguide layer, this cross-talk will not contribute to the side lobes. This is because the phase difference between each layer is zero, and the intensity of the light in each layer are comparable, so the vertical cross-talk in the whole system is in a dynamic equilibrium: when the main light pulse in one waveguide induces a delayed pulse in the adjacent waveguide, this waveguide will also receive a delayed pulse induced by the main pulse from that adjacent waveguide, and since the main pulse in each waveguide have zero phase difference, the induced delayed pulse in these waveguides also have zero phase difference. In this case, all the induced pulse can also interfere with each other in the same direction with the out-coupling beam, so it also contributes to the main lobe. Another simulation is done to confirm this, the result is shown in Fig. 5. In this simulation, we use 8 $Si_3N_4$ layers of 650nm, and 7 $SiO_2$ layers of 300nm, so the OPA still covers a $7.3\mu m$ range vertically, which is the same with the original structure in Fig. 1.

The horizontal angle steering range shown in Fig. 5 (a) is 23.00°/100nm, this is slightly different from the result in Fig. 4, the reason for this difference is the waveguide thickness is changed in this structure, so it will change the effective index of the waveguide; however, the pattern in this structure is the same as the previous one, so the change is not large. Fig. 5 (b, c) shows the vertical angle steering of this structure, it can be found, even though the $SiO_2$ layers are only 300nm thick, the vertical angle of the beam is not influenced by the cross-talk. However, the vertical FWHM and efficiency become larger, this is because that even though the OPA covers the same range vertically, the proportion of waveguide layers is higher, so compared to the original structure used for Fig. 4, this structure is more close to a thick slab waveguide with thickness of $7.3\mu m$. While in either case, the results in Fig. 5 is a clear evidence to show that the vertical cross-talk between waveguides does not influence the angle of emitting beam.

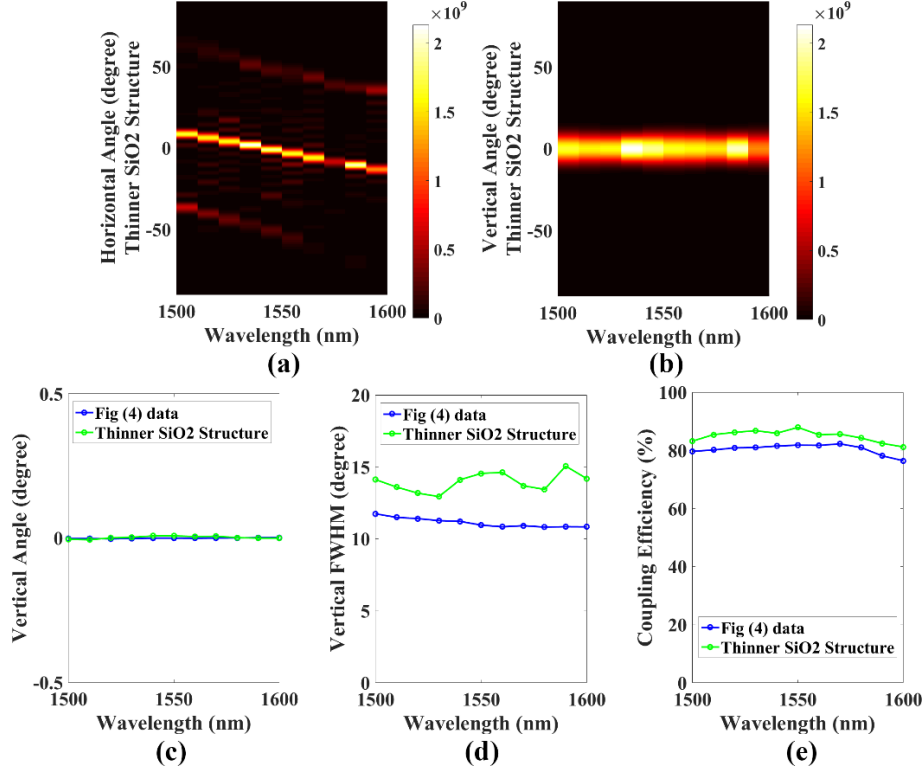

Fig. 5. Simulation result of the thinner SiO$_2$ structure (8 Si$_3$N$_4$ layers with 650nm thickness and 7 SiO$_2$ layers with 300nm thickness). (a) Horizontal farfield contour map, (b) Vertical farfield contour map, (c) Comparison of vertical angle between the thinner SiO$_2$ structure and the original structure, (d) Comparison of vertical FWHM, (e) Comparison of coupling efficiency.

### 3.2 Engineering of the delay length

We have pointed out that the horizontal convergence of the device can be further enhanced by using more waveguides in each layer. So, in the real case, the detection resolution of a wavelength tuned Lidar depends on the steering sensitivity per wavelength and the wavelength tuning resolution of the light source. In this work, we select 6200nm as the delay length of the structure; while in real application, the delay length can be selected larger to increase the steering sensitivity. Two simulations with delay length of 5400nm and 7000nm are done, the results are shown in Fig. 6.

The difference between the structures in this simulation is in the pattern of each waveguide layer, so only the information about the horizontal angle is plotted in Fig. 6. The most important comparison is in (c), in this figure, the red curve is the same red curve in Fig. 4 (c), the yellow curve is for the structure with 5400nm delay length and the pink curve is for 7000nm delay length. The steering sensitivity of 5400nm delay length structure is $21.58°/100$nm wavelength, is $24.78°/100$nm for the original device with delay length of 6200nm and is $28.32°/100$nm for delay length of 7000nm. This indicates that the steering sensitivity can be changed by selecting different delay length: in most cases, a larger delay length will be preferred to achieve higher steering sensitivity; while in some cases, if the wavelength tuning resolution of the light source is limited, it may require a lower steering sensitivity to increase the detection resolution, this can be done by selecting a smaller delay length. In the meantime, the horizontal FWHM and coupling efficiency do not change much, this is because the spacing between each array element is kept in all the three structures.

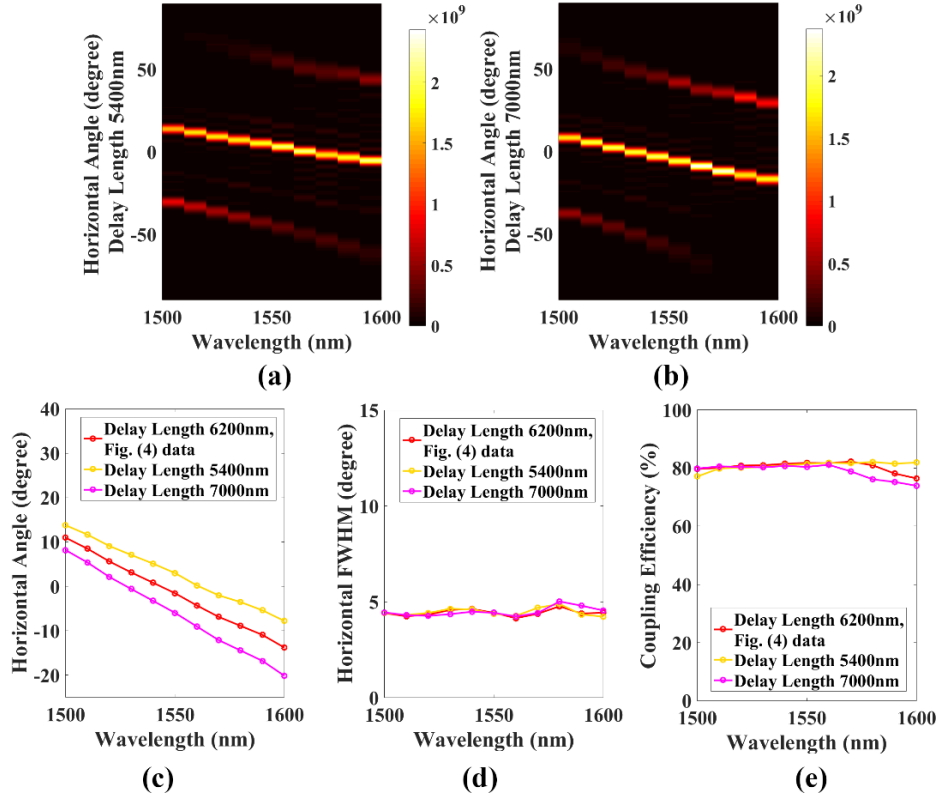

Fig. 6. Simulation result of structures with different delay length (5400nm and 7000nm). (a) Horizontal farfield contour map of the structure with delay length of 5400nm, (b) Horizontal farfield contour map of the structure with delay length of 7000nm, (c) Comparison of horizontal angle between the structures with different delay length (original 6200nm, 5400nm, 7000nm), (d) Comparison of horizontal FWHM, (e) Comparison of coupling efficiency.

### 3.3 Selection of the number of waveguide layers

In this work, we select 6 $Si_3N_4$ layers to cover a range of $7.3\mu m$ in vertical direction, this is to ensure the total vertical size is not larger than the mode field diameter (MFD) of a common single mode fiber. In this structure, the number of waveguides in each layer can be increased by the beam splitter tree [1], while the number of the waveguide layers is limited by the MFD of the field. On the other hand, as discussed in Sec. 2, the fabrication of multi-layer structure will become a challenge when more layers are required. So, the selection of how many layers to fabricate will be a tradeoff between the fabrication complexity and the device performance. In this work, we also investigate this parameter. Two structures which are exactly the same as the structure used for Fig. 4 but only different in the number of waveguide layers are simulated, the results are shown in Fig. 7.

In Fig. 7, only the information about the vertical angle is plotted, and similar to Fig. 6, the blue curve in (c-e) is exactly the same as the blue and black curve in Fig. 4 (c-e). As discussed before, the phase difference between each array element in vertical direction is always 0, so it is not a surprise that both the 4-layer structure and 8-layer structure shows similar farfield steering curve as the original 6-layer structure. However, in Fig. 7 (d), it can be found that the 6-layer structure actually shows the best FWHM result: the 4-layer structure shows a much larger FWHM because of the lack of enough array elements and vertical size; and the 8-layer structure also shows a similar but slightly higher FWHM. The reason that 8-layer structure doesn't show a better FWHM may because of the vertical cross-talk: it is not an issue in less layer structure, but may become an issue in more layer structure. In Fig. 7 (c), the vertical angle of the 8-layer structure still doesn't change much in the whole wavelength range, but the whole curve shifts a little bit to positive, this may also come from the vertical cross-talk. This suggests that 6 layers may already be enough in this structure configuration. On the other hand, the 4-layer structure also shows a clear convergence, even though the FWHM is wider than the 6-layer structure. This may be helpful if there are some applications where the vertical convergence is not critical, then the 4-layer structure will significantly reduce the fabrication complexity.

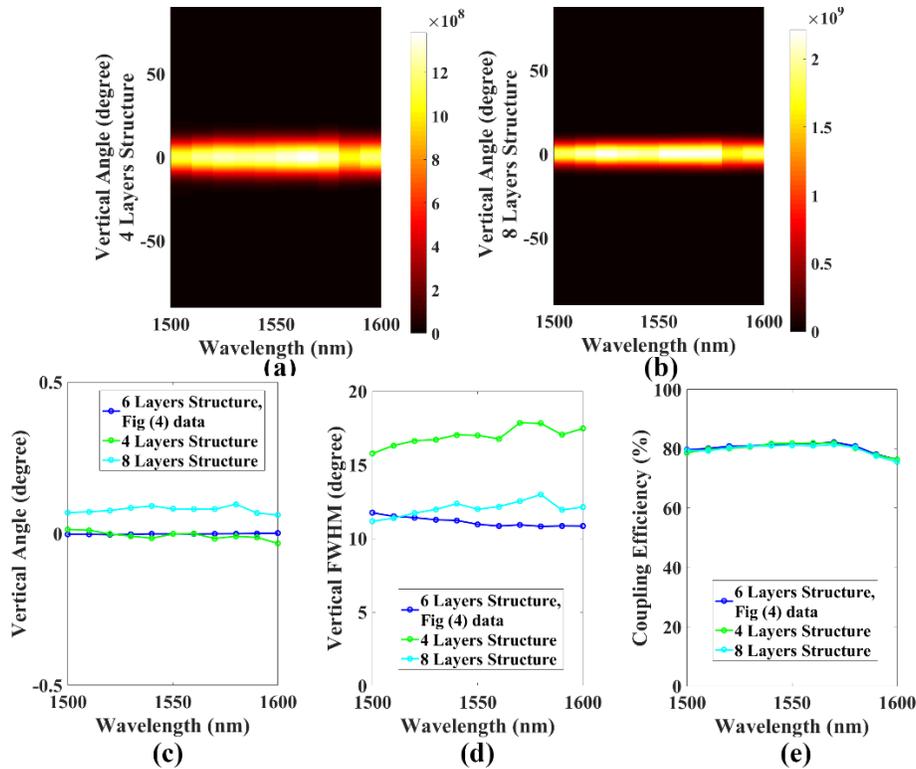

Fig. 7. Simulation result of structures with different waveguide layers (4 layers and 8 layers). (a) Vertical farfield contour map of the structure with 4 waveguide layers, (b) Vertical farfield contour map of the structure with 8 waveguide layers, (c) Comparison of vertical angle between the structures with different waveguide layers (original 6 layers, 4 layers, 8 layers), (d) Comparison of vertical FWHM, (e) Comparison of coupling efficiency.

## 4. Conclusion

In this work, a 3-D optical phased array (OPA) with the light exiting from the edge of the device, which is based on multi-layer $Si_3N_4/SiO_2$ platform, is numerically demonstrated. The CMOS compatible fabrication strategy of this device is discussed. The multi-layer structure can enable a high efficiency from both the input coupling and emitting coupling, the end-fire emitting efficiency can be as high as 82%. A 2-D converged beam is clearly generated in the farfield pattern of the emitting OPA, which can be steered purely horizontally by wavelength tuning, this suggests a possibility to apply the device to build a multi-line solid state Lidar. The inter-relationship of the 3-D OPA structure is studied in detail: the vertical cross-talk doesn't affect the out-coupling angle; the length of delay line can be engineered to achieve either high steering sensitivity or high steering resolution; the number of waveguide layers can also be engineered as a trade-off between the fabrication complexity and device performance. The two main features, high efficiency and single degree of freedom control, are explained in detail. This work is promising for further study of solid-state beam steering devices and the application of solid-state Lidar, as well as in other emerging areas, such as wireless communication or optical microscope.

**Funding.** National Science Foundation (NSF) (NSF-1710885); University of Michigan (U-M); Toyota Research Institute at North America.

**Disclosure.** F: Toyota Research Institute at North America.